\documentclass[english,12pt,aps,prd,a4paper,preprintnumbers,floatfix,nofootinbib,showpacs,superscriptaddress, notitlepage]{revtex4-1} 
\usepackage[mode=buildnew]{standalone}
\usepackage[usenames,dvipsnames]{color}  
\usepackage{graphicx}

\usepackage{setspace}
\usepackage{caption}
\usepackage{subcaption}
\usepackage{amsmath}
\usepackage{amssymb}
\usepackage[colorlinks=true,citecolor=darkred,urlcolor=darkred, pdfborder={0 0 0}]{hyperref}
\usepackage[normalem]{ulem}
\usepackage{xcolor}
\usepackage{placeins}
\usepackage{mathrsfs}
\usepackage{verbatim} 
\usepackage{cancel} 
\usepackage{float} 

\usepackage{scalerel}
\usepackage{tikz-feynman}
\tikzfeynmanset{compat=1.1.0}
\usepackage{feynmp}
\usepackage{tikzsymbols}
\usepackage{multirow}

\makeatletter
\def\p@subsection{}
\makeatother

\definecolor{darkred}{rgb}{0.6,0,0}

\definecolor{linkcolor}{rgb}{0,0,0.5}

\def\gsim{\raise0.3ex\hbox{$\;>$\kern-0.75em\raise-1.1ex\hbox{$\sim\;$}}}
\def\lsim{\raise0.3ex\hbox{$\;<$\kern-0.75em\raise-1.1ex\hbox{$\sim\;$}}}

\def\beqn#1{\begin{equation}\label{#1}}
\def\eeqn{\end{equation}}

\def\beqa#1{\begin{eqnarray}\label{#1}}
\def\eeqa{\end{eqnarray}}

\newcommand {\ignore}[1]{}

\def\Z4{$Z_4$}
\def\O5{$\mathcal{O}_5$ }

\def\321{$\mathrm{SU(3) \otimes SU(2) \otimes U(1)}$ }

\usepackage{booktabs}
\usepackage{slashed}

\begin{document}

\title{\boldmath \color{Blue} Generalized Chiral \(U(1)_{B-L}\) with Inelastic Scalar Dark Matter for the LZ 248 keV Event}

 \author{Ranjeet Kumar}\email{kumarranjeet.drk@gmail.com}
 \affiliation{Institute for Convergence of Basic Studies, Seoul National University of Science and Technology, Seoul 01811, Republic of Korea}

\author{Hemant Kumar Prajapati}
\email{hemant19@iiserb.ac.in}
\affiliation{Department of Physics, Indian Institute of Science Education and Research - Bhopal \\ Bhopal ByPass Road, Bhauri, Bhopal 462066, India}

\begin{abstract}
\noindent
The high-energy nuclear recoil event with \(E_R\simeq248\) keV recently reported by the LZ collaboration, with a background-only significance of \(2.6\sigma\), offers an intriguing window into dark matter scattering beyond the conventional elastic picture. We study an inelastic scalar dark matter scenario within a generalized chiral \(U(1)_{B-L}\) framework.
After symmetry breaking, the dark sector scalar gives rise to two nearly degenerate states with a small mass splitting. The lighter state serves as the dark matter candidate and couples off-diagonally to the \(Z'\) boson, leading to endothermic inelastic scattering. The resulting kinematics suppress the contribution from low-velocity dark matter while making the high-velocity tail increasingly relevant for nuclear recoils at higher energies. We identify the parameter space consistent with the observed relic abundance and current experimental constraints, and discuss the inelastic scattering kinematics relevant to the recent LZ observation, together with the complementary collider prospects for the associated \(Z'\) boson.

\end{abstract}

\maketitle

\section{Introduction} \label{sec:intro}

The existence of dark matter (DM) is firmly established by a wide range of cosmological and astrophysical observations, while its nature remains unknown \cite{Zwicky:1933gu,Rubin:1970zza,Rubin:1980zd}. Among the various possibilities, thermal weakly interacting massive particles (WIMPs)~\cite{Lee:1977ua,Kolb:1990vq,Jungman:1995df} provide one of the simplest and most extensively studied mechanisms for producing the observed relic abundance \cite{Planck:2018vyg}. The WIMP framework is particularly attractive because the same interactions responsible for thermal freeze-out can, in principle, lead to observable signals in direct detection and collider experiments.
Despite extensive experimental efforts, conventional elastic WIMP scattering has not yet been established. The resulting stringent
constraints on the DM-nucleon cross section have
motivated scenarios in which the scattering kinematics differs from the standard elastic case. 

In particular, inelastic DM provides a well-motivated possibility in which the incoming DM particle scatters into a slightly heavier state,
\begin{equation}
    \chi_1 + N \rightarrow \chi_2 + N,
    \qquad
    \delta \equiv m_{\chi_2}-m_{\chi_1}>0.
\end{equation}
The positive mass splitting introduces an endothermic threshold and
therefore modifies the minimum velocity required for a given nuclear
recoil. As a consequence, the low velocity bulk of the Galactic DM distribution can be strongly suppressed, while the high velocity tail can dominate the observable recoil spectrum.
This possibility has recently attracted renewed interest following the high-energy nuclear-recoil event reported by the LUX-ZEPLIN (LZ) collaboration~\cite{LZ2026HighRecoil}. The event, with recoil energy around $E_R\simeq 248$~keV, lies substantially above the recoil energy region usually associated with conventional elastic WIMP searches. Although
the statistical significance of the event is insufficient to establish a DM signal, its unusual recoil energy provides a well-defined
motivation for exploring non-standard scattering kinematics. In
particular, several recent studies have investigated whether an endothermic inelastic transition can account for such a high-energy recoil~\cite{Freese:2026sga,Su:2026rwz,Yin:2026jnn,Du:2026guj,Wu:2026nhi,Fan:2026kxx,Lou:2026idn,Visinelli:2026kgt,Yamashita:2026ump,DiMauro:2026ldr,Nomura:2026qyq,Smirnov:2026aqk,Unwin:2026rdp,Jeesun:2026vzo,Rodd:2026tyn,Chattopadhyay:2026ryw,Dent:2026bji,Gu:2026vto,deLima:2026shq,Borah:2026zwf,Bandyopadhyay:2026gjw,Das:2026uyy,Bose:2026ndd,Baer:2026fpy,Lee:2026wof,Okada:2026eol,cheung2026luxzeplincollidersprobinghiggsino}.
The LZ observation therefore motivates a broader investigation of inelastic DM scenarios in which a small mass splitting can substantially modify the recoil spectrum and enhance the high-energy region.

In this work, we consider a scalar realization of this inelastic framework.
In models with an extra Abelian gauge symmetry, a complex scalar DM can be decomposed into two real components after symmetry breaking. If the scalar potential contains a gauge-invariant trilinear term
coupling the scalar DM to the $U(1)$ breaking Higgs field, the two real components acquire a small mass splitting, and the lighter of them becomes the stable DM candidate \cite{Okada:2019sbb,Okada:2026eol}.
The associated gauge interaction is then off-diagonal in the real scalar basis. The resulting $Z'$ mediated scattering
\begin{equation}
    \chi_1 N\rightarrow \chi_2 N,
\end{equation}
is therefore intrinsically inelastic. This structure has important consequences for both the thermal history and direct detection. During freeze-out, the nearly degenerate scalar states can both contribute to the effective annihilation rate through co-annihilation, while the small mass splitting renders the $Z'$-mediated nuclear scattering endothermic.
Importantly, the direct detection rate is then controlled not only by
the nominal interaction strength but also by the kinematic threshold
associated with the mass splitting.

The above features can be naturally realized within the generalized chiral $U(1)_X$ framework developed in our recent work \cite{Kumar:2026rul}, where the anomaly-free fermion charges are parametrized by two independent parameters. A suitable charge choice suppresses the $Z'$ branching fraction into charged leptons, thereby weakening constraints from high-mass dilepton searches at the LHC. The framework also accommodates neutrino masses and provides a stable singlet scalar DM candidate.
In the present work, we consider a framework where these generalized chiral $U(1)_X$ charges allow a trilinear gauge invariant term in the scalar potential to generate a small mass splitting between the real and imaginary components of the singlet scalar DM.
The resulting dark sector contains a stable lighter state and a slightly heavier partner, with an off-diagonal \(Z'\) interaction that mediates the inelastic scattering process \(\chi_1 N\rightarrow\chi_2 N\).
We investigate the resulting DM phenomenology, focusing on the relic abundance, inelastic scattering, and the associated \(Z'\) constraints. In particular, we explore the parameter space consistent with the observed relic abundance and discuss the inelastic scattering kinematics relevant to the recent LZ observation, together with the complementary prospects of \(Z'\) searches at colliders.

The paper is organized as follows. In Sec.~\ref{sec:mod} we briefly review the
generalized chiral $U(1)_X$ framework and introduce the modified dark
sector. In Sec.~\ref{sec:inelst} we discuss the inelastic $Z'$ interaction and corresponding direct detection cross section. The numerical analysis is presented in Sec.~\ref{sec:num}. Finally, we summarize our results in Sec.~\ref{sec:conc}.

\section{Generalized chiral $U(1)_X$ framework} \label{sec:mod}

In the generalized framework \cite{Prajapati:2024wuu,Kumar:2026rul}, we consider an extension of the Standard Model (SM) by an additional Abelian gauge symmetry, \(U(1)_X\), with three right handed neutrinos, $\nu_{R_i}$ ($i=1,2,3$). The generalized fermion charge assignment is chosen to satisfy all gauge and mixed gravitational anomaly cancellation conditions as well as the gauge invariance of the SM Yukawa interactions.
Following our previous construction, the charges of the SM fermions and right handed neutrinos can be expressed in terms of two independent
parameters, which we denote by $X_L$ and $\kappa$, as shown in Table. \ref{tab:Charges}. 
\begin{table}[!h]
\centering
\renewcommand{\arraystretch}{1.5}
\setlength{\tabcolsep}{10pt}

\begin{adjustbox}{max width=\textwidth}
\begin{tabular}{|c|c|c|c|c|c|c|c|c|}
\hline
$Q$ &
$u_{R}$ &
$d_{R}$ &
$L$ &
$e_{R}$ &
$\nu_{R_1}$ &
$\nu_{R_2}$ &
$\nu_{R_3}$ &
$H$ \\
\hline\hline
$\displaystyle \frac{X_{L}}{3}$ &
$\displaystyle \frac{4X_{L}}{3}-\kappa$ &
$\displaystyle \kappa-\frac{2X_{L}}{3}$ &
$\displaystyle -X_{L}$ &
$\displaystyle \kappa-2 X_{L}$ &
$\displaystyle -4\kappa$ &
$\displaystyle -4\kappa$ &
$\displaystyle  5\kappa$ &
$\displaystyle X_{L}-\kappa$ \\
\hline
\end{tabular}
\end{adjustbox}
\caption{Anomaly-free \(U(1)_X\) charge assignments for the SM fermions, \(\nu_{R_i}\), and \(H\) in the generalized chiral \(U(1)_{B-L}\) framework, parametrized by \(\kappa\) and \(X_L\). The lepton doublet charge is chosen as \(-X_L\).}
\label{tab:Charges}
\end{table}
The right handed neutrinos carry the non-universal charges
\begin{equation}
    X_{\nu_{R_i}}=(-4\kappa,-4\kappa,5\kappa).
\end{equation}
The conventional chiral $B-L$ realization is recovered for the choice $\kappa = X_{L} =1$.

The resulting generalized charge structure has an important consequence
for the phenomenology of the new neutral gauge boson $Z'$. In
particular, the enlarged invisible sector can reduce the branching
fraction into charged leptons, thereby weakening the strongest
constraints from high-mass dilepton resonance searches \cite{ATLAS:2019erb}.
The $U(1)_X$ symmetry is spontaneously broken by the vacuum expectation value of an SM singlet scalar, $\eta$, carrying $q_{\eta}$ charge under $U(1)_X$,
\begin{equation}
    \eta =
    \frac{1}{\sqrt{2}}\left(v_\eta+\rho_\eta+iG_\eta\right),
\end{equation}
where $G_{\eta}$ is the Goldstone boson associated with the massive $Z'$.
The scalar and fermion sectors responsible for neutrino mass generation
are inherited from the generalized framework discussed in Ref. \cite{Kumar:2026rul} and will not be repeated
in detail here. Our focus is the modification of the dark sector and its consequences for inelastic scattering.

\subsection{Dark sector charge assignment}

We introduce a complex SM singlet scalar $\chi_d$ as the dark sector
field with the following charge assignment under SM$\otimes U(1)_X$ symmetry,
\begin{equation}
    \chi_d \sim (1,1,0,q_d).
\end{equation}
In the previous realization, the charge of $\chi_d$ was chosen such
that the field remained stable but its real and imaginary components
were effectively degenerate. Here we modify the charge assignment to
allow a symmetry breaking interaction of the form
\begin{equation}
    V_{\rm split}
    =
    \frac{\mu_d}{2}
    \left(\chi_d^2\eta^\ast+\mathrm{H.c.}\right),
\end{equation}
where gauge invariance requires
\begin{equation}
    2q_d-q_\eta=0.
\end{equation}
Thus,
\begin{equation}
    q_d=\frac{q_\eta}{2}.
\end{equation}

This interaction is the essential new ingredient of the present
construction. After $U(1)_X$ breaking, it generates a small splitting
between the real and imaginary components of $\chi_d$.
We decompose the dark scalar as
\begin{equation} \label{eq:dm}
    \chi_d =
    \frac{1}{\sqrt{2}}(\chi_1+i\chi_2),
\end{equation}
where $\chi_1$ denotes the lighter state and $\chi_2$ the heavier state.
The relevant scalar potential can be written as
\begin{align} \label{eq:pot}
- V \supset\,
m_d^2|\chi_d|^2
+\lambda_{Hd}|H|^2|\chi_d|^2
+\lambda_{\eta d}|\eta|^2|\chi_d|^2 - 
\left[
\frac{\mu_d}{2}\chi_d^2\eta^\ast
+\mathrm{H.c.}
\right],
\end{align}
After symmetry breaking, the scalar fields $H$ and $\eta$ can be written as,
\begin{equation} \label{eq:exp}
    H = 
    \begin{pmatrix}
    G^+\\
    (v_H+h+i G_H)/\sqrt{2}
    \end{pmatrix},
    \qquad
    \eta =
    \frac{1}{\sqrt{2}}\left(v_\eta+\rho_\eta+iG_\eta\right).
\end{equation}
Utilizing Eqs. \eqref{eq:dm}-\eqref{eq:exp}, the masses of the two dark sector scalar can be computed as
\begin{align}
    m_{\chi_1}^2 &=
    m_d^2
    +\frac{1}{2}\lambda_{Hd}v_H^2
    +\frac{1}{2}\lambda_{\eta d}v_\eta^2
    -\frac{\mu_d v_\eta}{\sqrt{2}},
    \\
    m_{\chi_2}^2 &=
    m_d^2
    +\frac{1}{2}\lambda_{Hd}v_H^2
    +\frac{1}{2}\lambda_{\eta d}v_\eta^2
    +\frac{\mu_d v_\eta}{\sqrt{2}}.
\end{align}
Consequently, we get
\begin{equation}
    m_{\chi_2}^2-m_{\chi_1}^2
    =
    \sqrt{2}\,\mu_d v_\eta.
\end{equation}
For a small splitting, we can approximate the mass splitting parameter $\delta$ as,
\begin{equation}
    \delta \equiv m_{\chi_2}-m_{\chi_1}
    \simeq
    \frac{\mu_d v_\eta}{\sqrt{2}m_{\chi_1}}.
\end{equation}
Thus, a sub-MeV splitting can naturally arise from a small parameter $\mu_d$. 

\section{INELASTIC interaction and direct detection} \label{sec:inelst}

The dark sector scalars couple to the \(Z'\) gauge boson through an off-diagonal interaction of the form
\begin{equation}
    {\cal L}_{\rm int}
    =
   q_{d} g_x Z'_\mu
    \left[
    \chi_2\partial^\mu \chi_1-\chi_1\partial^\mu \chi_2
    \right].
\end{equation}
Importantly, there is no diagonal $Z'\chi_1 \chi_1$ or $Z'\chi_2 \chi_2$ interaction.
Therefore, the leading $Z'$-mediated nuclear scattering process is
\begin{equation} \label{eq:inels}
    \chi_1+N\rightarrow \chi_2+N.
\end{equation}
This is the central feature of the present model. The same gauge interaction that controls the dark sector thermal history can therefore lead to an intrinsically inelastic direct detection signal.
The stability of the lighter state is ensured by the residual symmetry
left after $U(1)_X$ breaking \cite{Bonilla:2018ynb,CentellesChulia:2019gic,Srivastava:2019xhh,Kumar:2025cte,Kang:2026osw,Kang:2026lgr}, provided the complete charge assignment for the dark sector forbids operators inducing its decay. We impose
this condition throughout our analysis.

%
For a nuclear recoil energy $E_R$, the minimum velocity required for
the inelastic transition provided in Eq.~\eqref{eq:inels} is given by,
\begin{equation}
    v_{\rm min}(E_R)
    =
    \frac{1}{\sqrt{2m_N E_R}}
    \left(
    \frac{m_N E_R}{\mu_{\chi N}}
    +\delta
    \right),
\end{equation}
where
\begin{equation}
    \mu_{\chi N}
    =
    \frac{m_{\chi_1}m_N}{m_{\chi_1}+m_N}
\end{equation}
is the DM-nucleon reduced mass.
For elastic scattering, $\delta=0$, and the minimum velocity is
determined entirely by the recoil energy. In the present case, the
additional positive term proportional to $\delta$ raises the velocity
threshold.
This has two important consequences. First, scattering from the low velocity part of the Galactic DM population is suppressed.
Second, the recoil spectrum is shifted towards higher recoil energies.
Consequently, a relatively large nominal scattering cross section does not translate directly into the standard elastic DM exclusion
limits.
This distinction is essential when comparing the present model with conventional direct detection constraints.


\section{Numerical analysis} \label{sec:num}

Up to this point the construction has been kept general, with the
$U(1)_X$ charges parametrized by the two free quantities $X_L$ and
$\kappa$. The inelastic $Z'$-mediated nuclear scattering of
Eq.~\eqref{eq:inels} is a generic feature of the framework and is
obtained for any anomaly-free choice of these parameters, since it
follows solely from the requirement $q_d=q_\eta/2$.
%
For the phenomenological analysis, however, we must fix a definite
benchmark. We adopt the simplest choice, $X_L=\kappa=1$, for which the SM fermion charges coincide with those of the conventional $(B-L)$ assignment and the right handed neutrinos carry the non-universal
charges $(-4,-4,5)$.
While we have  $q_\eta= 3$ and $q_d=q_\eta/2=3/2$ for the scalars $\eta$ and $\chi_d$, respectively.
Since the SM Higgs doublet is uncharged under $U(1)_X$, no tree-level
$Z-Z'$ mass mixing is generated and the $Z'$ mass is given as, $M_{Z'} =3v_{\eta}g_{x}$.
Furthermore, the quark charges under $U(1)_{B-L}$ are vectorial, so
that their couplings to the $Z'$ are vector like as well.
The resulting DM-nucleon scattering is therefore spin independent (SI).
The corresponding cross section between the DM candidate and a nucleon $N$ can
then be expressed as
\begin{equation}
    \sigma_{\rm SI}
    \simeq
    \frac{9\mu_{\chi N}^2}{4\pi}
    \left(
    \frac{g_x^2}{M_{Z'}^2}
    \right)^{\!2},
\end{equation}
where $\mu_{\chi N}=m_{\chi_1}m_N/(m_{\chi_1}+m_N)$ is the DM-nucleon
reduced mass and we have used the approximation $m_{\chi_1}\simeq
m_{\chi_2}$.
Notice that for DM of order TeV one has $\mu_{\chi N}\simeq m_N$, and hence the magnitude of the cross section is a function of only $g_x$ and $M_{Z'}$. 
On the other hand, the mass splitting
$\delta = m_{\chi_2}-m_{\chi_1}$, of a few hundred keV, is important
and sets a kinematic threshold that suppresses standard low-energy
recoils, shifting the peak of the differential event rate to higher
recoil energies, as observed at LZ.

An inelastic scattering cross section of order
$\mathcal{O}(10^{-45})\,\text{cm}^{2}$, for a mass splitting
$\sim \mathcal{O}(100)$~keV, is required to account for the LZ anomaly \cite{LZ2026HighRecoil,Okada:2026eol}.
We adopt the benchmark point (BP) corresponding to values $M_{Z'}=5$~TeV and $g_x=0.3$ for the
$Z'$ mass and gauge coupling, which yields
$\sigma_{\rm SI} \approx 3\times10^{-45}\,\text{cm}^{2}$, within the range favored by the LZ event.
This benchmark also evades the constraints from LEP-II, which requires
$M_{Z'}\geq 6.9\,g_x$~TeV \cite{Electroweak:2003ram,Carena:2004xs,ALEPH:2013dgf}, as well as those from the high-$p_T$
Drell-Yan tails, $M_{Z'}\geq 9.2\,g_x$~TeV \cite{Allwicher:2022gkm,Allwicher:2022mcg}.  
To demonstrate that this benchmark is also compatible with the observed
relic density \cite{Planck:2018vyg} and with LHC constraints from dilepton resonance
searches \cite{ATLAS:2019erb}, we implement the model in \textit{SARAH}~\cite{Staub:2015kfa}
to generate the particle spectrum, interaction vertices, and the model
files required for the numerical analysis. These are interfaced with
\textit{SPheno}~\cite{Porod:2003um} to obtain the mass
spectrum and the relevant decay widths, and the resulting output is
passed to \textit{MadGraph5}~\cite{Alwall:2014hca}, where the
production cross sections are evaluated at leading order for
proton-proton collisions at $\sqrt{s}=13$~TeV. 
On the DM side, the SPheno spectrum is instead fed to
\textit{micrOMEGAs}~\cite{Belanger:2020gnr} to compute the relic
abundance.
 \begin{figure}[!h]
     \centering
     \includegraphics[width=0.46\linewidth]{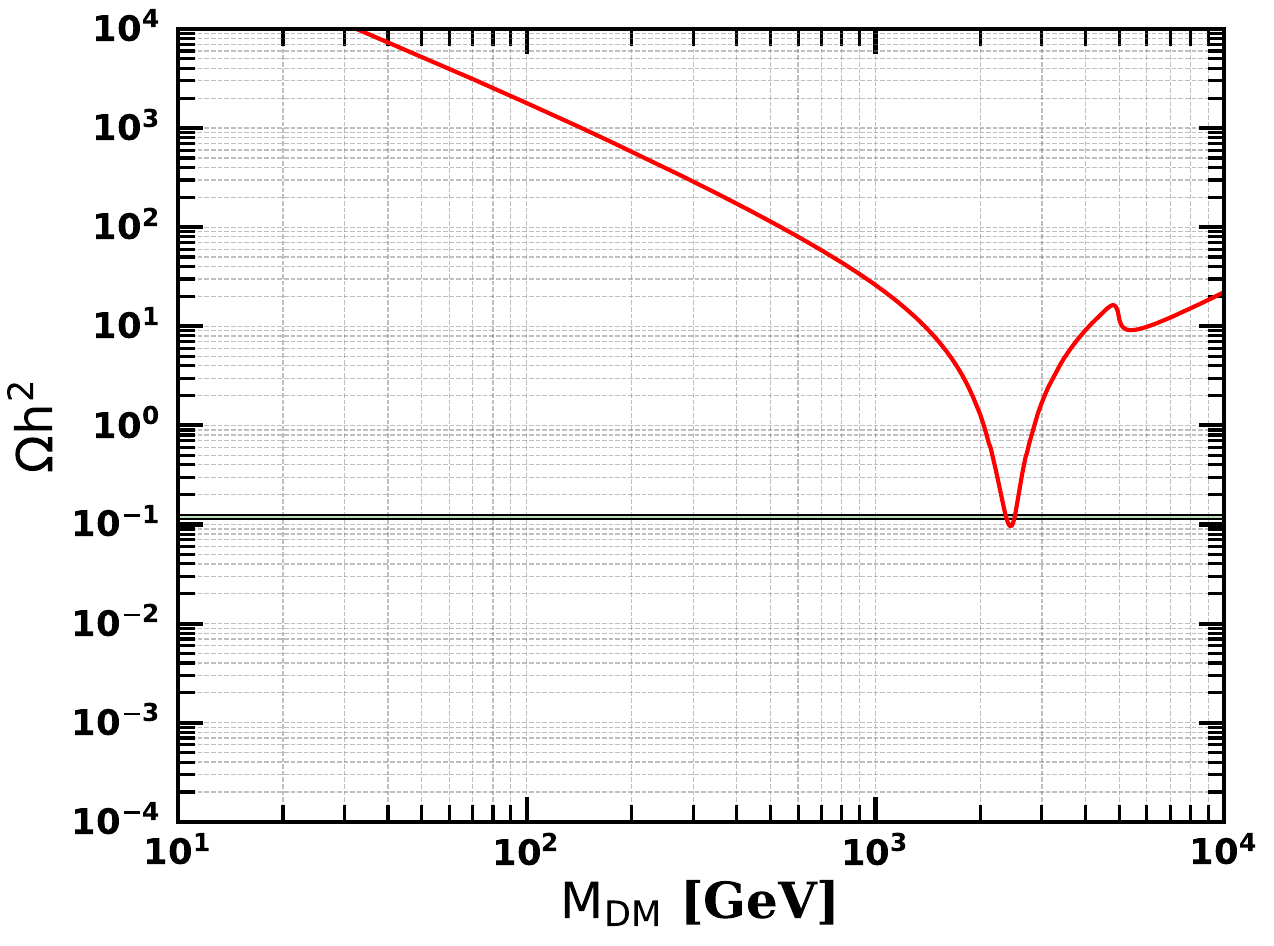}
      \includegraphics[width=0.49\linewidth]{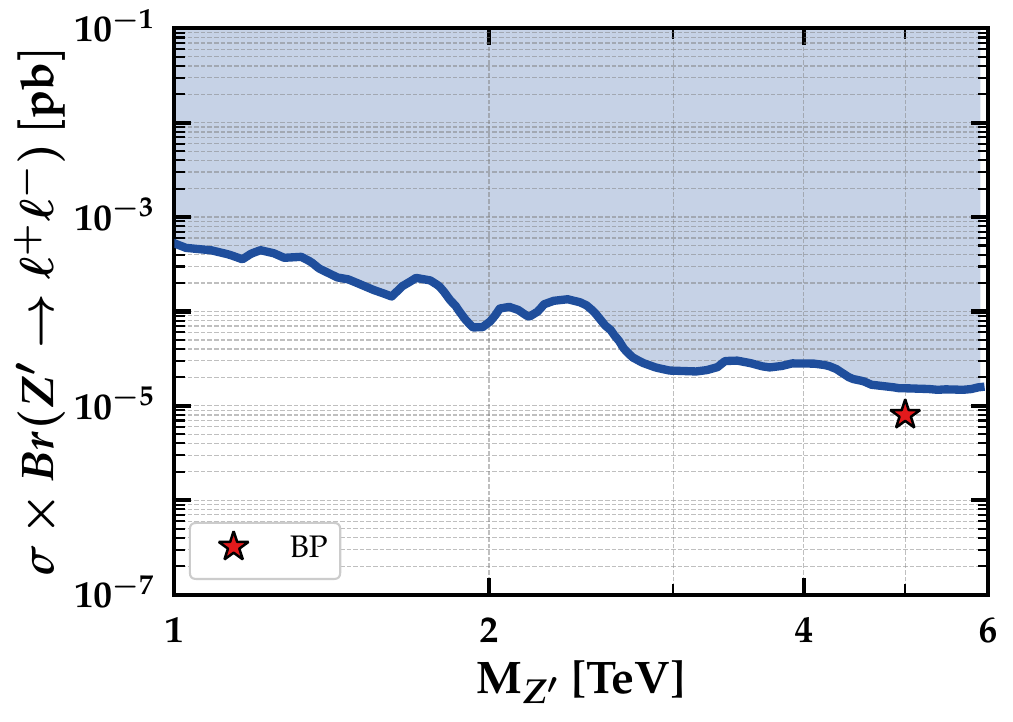}
    \caption{\textit{Left panel:} relic abundance $\Omega h^2$ as a
function of the DM mass $m_{\chi_1}$ for the benchmark
$M_{Z'}=5$~TeV and $g_x=0.3$. The horizontal band denotes the
Planck measured value \cite{Planck:2018vyg}. \textit{Right panel:} ATLAS limit on
$\sigma\times\mathrm{Br}(Z'\to\ell^+\ell^-)$ at $\sqrt{s}=13$~TeV
(blue), with the shaded region excluded. The star marks our
benchmark point (BP).}
     \label{fig:Limit}
 \end{figure}
Our results are summarized in Fig.~\ref{fig:Limit}. The left
panel shows the relic abundance as a function of the DM mass. A sharp
dip appears at $m_{\chi_1}\simeq M_{Z'}/2$, where the $s$-channel
$Z'$ exchange becomes resonant and the annihilation cross section is
strongly enhanced. The observed relic density is reproduced in this
region, which fixes the DM mass to $m_{\chi_1}\simeq2.5$~TeV for our
benchmark.
The right panel displays the corresponding collider constraint. Our
benchmark lies well below the ATLAS exclusion curve, since the
non-universal charges $(-4,-4,5)$ of the right handed neutrinos
substantially reduce $\mathrm{Br}(Z'\to\ell^+\ell^-)$ relative to the
minimal $(B-L)$ case. The benchmark is therefore simultaneously
consistent with the observed relic abundance, the dilepton resonance
searches, and the cross section required by the LZ event.



\FloatBarrier


\section{Conclusions} \label{sec:conc}

We have studied an inelastic scalar DM scenario within a generalized chiral \(U(1)_X\) framework, motivated by the high-energy nuclear recoil event reported by the LZ collaboration. A suitable charge assignment allows a symmetry breaking interaction in the dark sector to generate a small mass splitting between two real scalar states ($\chi_1$ and $\chi_2$). The lighter state $\chi_1$ serves as the DM candidate, while its off-diagonal coupling to the \(Z'\) leads to endothermic inelastic scattering.
The small mass splitting therefore has important consequences for the DM phenomenology. During freeze-out, the nearly degenerate states remain thermally populated, making co-annihilation relevant for obtaining the observed relic abundance. At the same time, the inelastic transition introduces a kinematic threshold in nuclear scattering, making the high-velocity tail of the DM distribution particularly relevant for high-energy recoils. We identify viable parameter space satisfying the relic density requirement and relevant experimental constraints, while also examining the corresponding phenomenology of the \(Z'\) sector.

The resulting scenario therefore connects the thermal history of scalar DM with inelastic scattering and the phenomenology of an additional \(Z'\) boson. The interplay among these sectors provides a viable framework for exploring DM parameter space in the presence of a small mass splitting, with direct detection and collider searches offering complementary probes.

\section*{Acknowledgments} 
\noindent
RK is supported by the National Research Foundation of Korea under Grant NRF-2023R1A2C100609111. The work of HKP is supported by the Prime Minister Research Fellowship (ID: 0401969). We would like to thank Rahul Srivastava for useful discussions and for his valuable comments on the manuscript.

\bibliographystyle{utphys}
\bibliography{references.bib} 
\end{document}